\documentclass[%
 amsmath,amssymb,
 aps,
 twocolumn,
 showpacs,
 superscriptaddress,
 groupedaddress,
 prb
]{revtex4}
\usepackage{amsmath,amssymb,bm}
\usepackage{hyperref}
\usepackage{graphicx}
\usepackage{epstopdf}
\usepackage{latexsym}
\usepackage[usenames, dvipsnames]{color}
\usepackage[usenames, dvipsnames]{xcolor}
\usepackage{braket}
\usepackage{float}
\usepackage[normalem]{ulem}
\usepackage{mathtools}
\usepackage{tabu}
\usepackage{multirow}
\usepackage[inline]{enumitem}
\usepackage{cleveref}
\usepackage{xcolor}
\usepackage[normalem]{ulem}
\usepackage{tabularx}
\usepackage{array}
\usepackage{makecell}
\usepackage{soul}
\usepackage{url}
\usepackage{amsfonts}
\usepackage{textcomp}
\usepackage{soul}
\def\BibTeX{{\rm B\kern-.05em{\sc i\kern-.025em b}\kern-.08em
    T\kern-.1667em\lower.7ex\hbox{E}\kern-.125emX}} 

\begin{document}

\title{
Majorana zero modes in quantum Hall edges survive edge reconstruction
}
\author{Kishore Iyer}
\affiliation{Aix Marseille Univ, Université de Toulon, CNRS, CPT, Marseille, France}
 
\author{Amulya Ratnakar}
\affiliation{Department of Physics, Indian Institute of Science Education and Research (IISER) Kolkata, Mohanpur - 741246, West Bengal, India}
\author{Sumathi Rao}
\affiliation{International Centre for Theoretical Sciences, Tata Institute
of Fundamental Research, Bengaluru 560089, India}
\author{Sourin Das}
\affiliation{Department of Physics, Indian Institute of Science Education and Research (IISER) Kolkata, Mohanpur - 741246, West Bengal, India}

\begin{abstract}
A smooth edge potential on a $\nu = 1$ quantum Hall system leads to a $\nu = 1/3$ side strip, which could yield both Majorana and parafermion zero modes at the domain wall of the superconductor and ferromagnet on the edge. However, constraints imposed by the $\nu = 1$ bulk allow only for a pair of Majoranas, leading to a $Z_2 \times Z_2$ symmetric ground state. Signatures of both Majoranas appear in the $4\pi$ fractional Josephson current when the edge velocities are taken to be different.
\end{abstract}

\maketitle
\emph{Introduction:--} Non-abelian anyons are particles that possess non-abelian braiding statistics, characterized by degenerate ground states, which can be leveraged to build fault-tolerant qubits \cite{non_abelian_rmp}. Experimental challenges impede the observation of non-abelian anyons in systems such as $\nu = 5/2$ fractional quantum Hall effect (FQHE) \cite{PhysRevLett.59.1776,Moore:1991ks} and $p+ip$ superconductors \cite{Kitaev_2001}, motivating efforts towards engineering non-abelian anyons in heterostructures of conventional materials \cite{Alicea_2012,doi:10.1146/annurev-conmatphys-030212-184337,doi:10.1146/annurev-conmatphys-030212-184337}. 

Recently, interest has risen in engineering parafermion zero modes (PZM) -- $Z_N$ generalization of Majorana zero modes (MZM) -- offering richer fault-tolerant qubit operations compared to Majorana-based qubits \cite{Fendley2012_parafermions, AliceaReview_parafermions}. Proposals to engineer parafermions involve a pair of FQH edges, proximitized alternatively by superconductors and insulators \cite{Clarke2013_parafermions,Lindner2012_parafermions,Cheng12,Vaezi13_parafermions,PhysRevX.4.031009,Mong14,udit&yuval_parafermion,Barkeshli14PRX,sau17}. Experiments in this direction have observed Crossed-Andreev reflection in graphene as well as 2DEG systems \cite{Lee2017_parafermion, gul2021andreev_parafermion,shabani22}, inspiring focused theoretical investigations in experimentally relevant regimes \cite{nielsen_dynamics_2023,nielsen_readout_2022,michelsen_current_2020,schiller_predicted_2020,schiller_superconductivity_2023,iyer23,snizhko18,tang_vortex-enabled_2022,flavio22,PhysRevResearch.4.043094}.

Edge reconstruction, a challenge in interpreting quantum Hall experiments, results from the interplay of electronic correlation in the FQH state and a smooth edge potential \cite{chamon_sharp_1994,wan_edge_2003,wan_reconstruction_2002}. Competition between the need to completely neutralize the positive background and that to form an incompressible droplet leads to the creation of  counter-propagating edge modes along the original edge. Edge renormalization refines the edge structure, potentially incorporating upstream neutral modes \cite{KFP, khanna_emergence_2022,khanna_edge_2022,khanna_fractional_2021}.

FQH edge experiments aimed at measuring topological properties of the FQHE rely on the bulk-edge correspondence \cite{tong2016lectures}. While the bulk invariant of the quantum Hall system remains intact under edge reconstruction, it destroys several aspects of the FQH bulk-edge correspondence, with potentially dramatic consequences \cite{mach_zehnder1,mach_zehnder2}. 

\begin{figure}
    \centering
    \includegraphics[scale=0.23]{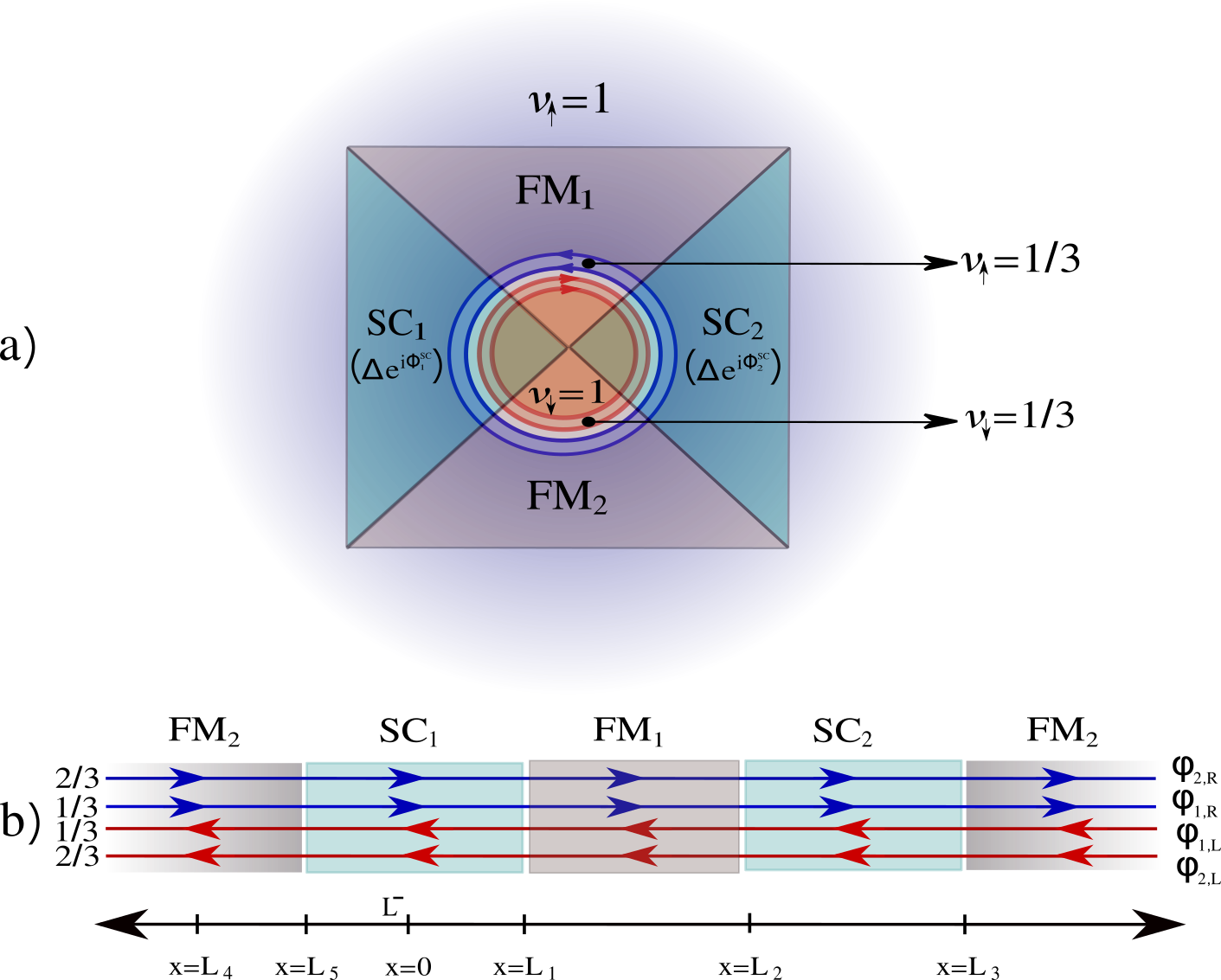}
    \caption{a) Displays two concentric $\nu_{\uparrow/\downarrow} =1$ quantum Hall systems, with the inner and outer systems having opposite spins. Edge reconstruction leads to the deposition of thin $\nu_{\uparrow/\downarrow} =1/3$ FQH bulks along the edges of the $\nu_{\uparrow/\downarrow} =1$ systems. Their edges are proximitized by alternating superconductors ($SC_{i}$) and ferromagnets ($FM_{i}$). $\phi_i^{SC}$ and $\Delta$ denote the phase and gap of the $SC_{i}$. b) Displays an unfolded version of our setup where the edge physics is transparent. Each reconstructed edge admits two bosonic modes, with an outer mode of conductance of 1/3$e^{2}/h$ ($\phi_{1R/L}$) and an inner mode with conductance of 2/3$e^{2}/h$ ($\phi_{2R/L}$). The neutral mode present in the reconstructed edges is not shown here. $R/L$ denotes the right/left moving bosonic edge modes.
 }
    \label{fig:set_up}
\end{figure}

Given the fundamental role of FQH edges in engineering parafermions, it is important to understand the consequences of edge reconstruction on PZM. As a first step, we study MZM built from reconstructed $\nu = 1$ quantum Hall edges, the main focus of this Letter. Edge reconstruction leads to the deposition of a $\nu = 1/3$ FQH side strip along the edge of the $\nu =1$ QH system, leading ultimately to two upstream bosonic charge modes of conductance 1/3($e^2/h$) and 2/3($e^2/h$). The edges of two such reconstructed $\nu =1$ QH systems are proximitized with alternating superconductors (SC) and ferromagnets (FM) as shown in Fig. \ref{fig:set_up}. The existence of several bosonic modes on the edge requires adding multiple pairing and backscattering terms to fully gap out the edges. \emph{A priori}, this leads to localized Majorana (due to $\nu =1$ bulk) and parafermion (due to $\nu =1/3$ bulk) zero modes on the domain walls between SC and FM. Consistency requirements between the multiple pairing and backscattering terms demote the PZM to an MZM, leading to two decoupled MZMs at each SC-FM domain wall. Hence, despite edge reconstruction, the periodicity of the Josephson current remains $4\pi$ though
signatures of the multiplicity due to the second Majorana appears in the fractional Josephson current when the propagation velocities of the two charged bosonic modes are different.

\emph{Review of $\nu=1$ edge reconstruction:--} In a $\nu = 1$ quantum Hall system, a sharp confining potential yields a single bosonic mode with conductance $e^2/h$ \cite{wen_theory_1992}. Smooth edge potentials can create a thin adjacent $\nu = 1/3$ fractional quantum Hall bulk, introducing two counter-propagating edge modes of conductance $1/3$. Edge renormalization, due to interactions and disorder-induced tunneling, results in an edge structure with downstream bosonic modes of conductance $2/3(e^2/h)$ and $1/3(e^2/h)$, and an upstream neutral mode \cite{khanna_fractional_2021, KFP}.

The bosonic Hamiltonian density of the reconstructed $\nu =1$ edge is given by $H_{R/L} = \sum_{\alpha=1}^{3}\frac{v_\alpha}{4\pi\nu_\alpha}(\partial_x\phi_{\alpha R/L}(x))^2$, where, $\phi_1$ and $\phi_2$ denote the bosonic modes with conductance $\nu_1 = 1/3$ and $\nu_2 = 2/3$ respectively. $\phi_3$ denotes the neutral mode (of zero conductance) with $\nu_3 = 2$. $R/L$ denotes the chirality and $v_\alpha$ denotes the velocity of $\phi_{\alpha}$. The bosonic fields $\phi_{\alpha R/L}$ obey the commutation relations $\left[ \phi_{\alpha R/L}(x),\phi_{\beta R/L}(y)\right] = \pm i\pi\nu_\alpha \delta_{\alpha\beta}\text{sgn}(x-y)$. 

Given this theory of the reconstructed $\nu = 1$ edge, all possible electron operators on the edge can be calculated \cite{shtanko14}. The most relevant electron operators on the considered $\nu = 1$ edge are found to be $\psi_A \sim e^{-i(\phi_1 + \phi_2)}$, $\psi_B \sim e^{-i3\phi_1}$ and $\psi_C \sim e^{-i3/2\phi_2-i/2\phi_3}$ \cite{[{See Supplemental Material at }][{ for details on theory of reconstructed edge, counting the degeneracy of the ground states, and calculation of the Josephson junction.}]supp}. $\psi_A$ is related to the electron operator on the unreconstructed $\nu =1$ edge, where the electronic charge is now split into a charge $e/3$ on $\phi_1$ and a charge $2e/3$ on $\phi_2$. $\psi_B$ and $\psi_C$ are made up solely of charged excitations on $\phi_1$ and, $\phi_2$ respectively.

The thin $\nu = 1/3$ bulk is taken to be large enough to accommodate fluxes. Let $n$ be the number of flux quanta in the $\nu = 1$ QH bulk, and $m$ that in the $\nu = 1/3$ FQH side strip. Bulk-edge correspondence dictates that the number of fluxes in the bulk is the same as the charge (quasi-particles) on the edge. The total charge on the reconstructed edge depends on both $\nu = 1$ and $\nu = 1/3$ FQH bulks. A flux quantum in the $\nu=1$ bulk induces a charge on both $1/3$ ($\phi_{1}$) and $2/3$ ($\phi_{2}$) bosonic modes (corresponding to an $\psi_A$ electron); however, a flux quantum in the $\nu=1/3$ bulk induces charge only on the $1/3$ ($\phi_{1}$) charge edge mode (three flux quanta correspond to a $\psi_B$ electron). The total charge on $\phi_j$, defined as $q_j = 1/2\pi\int dx~ \partial_x\phi_j$, with $q_1  = \frac{1}{3}(n + m)$ and
$q_2 = \frac{2}{3}n$. The total charge on the full edge is $q = q_1 + q_2 = n + m/3$, which in principle can be fractional. We will later see that as the edges are proximitized with superconductors and ferromagnets, energy considerations prevent a total fractional charge on the edges.

\emph{Theoretical model:--} We now bring together two such edge-reconstructed $\nu =1$ QH systems with opposite spins. Their edges are proximitized with alternating superconductors (SC) and ferromagnets (FM), as shown in Fig. \ref{fig:set_up}. The system is then described by the bosonized Hamiltonian, $H = H_{0} + H_{SC}+H_{FM}$ where
\begin{eqnarray}
    H_{0} &=&\sum_{\alpha=1}^{2}\frac{\hbar v_\alpha}{\nu_{\alpha}}\int dx \left[\left(\partial_{x}\varphi_{\alpha}(x)\right)^{2} + \left(\partial_{x}\theta_{\alpha}(x)\right)^{2}\right] \nonumber\\
    H_{SC} &=& -\int dx\lbrace\Delta_{A}\cos\left[2(\varphi_{1} + \varphi_{2}) \right] + \Delta_{B}\cos\left[6\varphi_{1} \right] \nonumber\\ && +\Delta_{C}\cos\left[3\varphi_{2} \right]\rbrace \nonumber\\
    H_{FM} 
     &=& -\int dx\lbrace\mathcal{M}_{A}\cos\left[2(\theta_{1} + \theta_{2})\right] + \mathcal{M}_{B}\cos\left[6\theta_{1}\right] \nonumber\\
    && + \mathcal{M}_{C}\cos\left[3\theta_{2}\right]\rbrace
    \label{Eq:pairing}
\end{eqnarray}
where $\nu_{\alpha}$ denotes the conductance of $\alpha^{th}$ edge, with $\nu_{1}=1/3$, $\nu_{2}=2/3$, and $v_\alpha$ denotes the velocity of the $\alpha$-th bosonic mode. $\Delta_\gamma $ and $\mathcal{
M_\gamma}$ are the SC and FM pairing amplitudes, having the spatial profile shown in Fig. \ref{fig:set_up} and $\varphi_\alpha/\theta_\alpha = (\phi_{\alpha R} \pm \phi_{\alpha L})/2 $. The neutral mode, which does not couple to the SC and FM, will be ignored. With the addition of the pairing and backscattering terms in Eq. \ref{Eq:pairing}, the edges are fully gapped, and the degeneracy of the ground state can be counted.

\emph{Ground-state manifold:--}
QH regions, whose edges are proximitized by an SC or an FM, can either exchange charge (with the SC) or spin (with the FM). Thus, these regions can be characterized by charge parity (in the SC region) or spin parity (in the FM region) operators. As explained earlier, by bulk-edge correspondence, these charges/spin are the same as the charges/spin at the edges, with them being distributed among the different bosonic modes as
\begin{equation}
\begin{split}
    \hat{Q}^{(\alpha)}_{i}
    &=  \frac{1}{\pi}\int_{SC_{i}} dx~ \partial_{x} \theta_{\alpha}(x)  \\
    \hat{S}^{(\alpha)}_{i} 
    &=   \frac{1}{\pi}\int_{FM_{i}} dx~ \partial_{x} \varphi_{\alpha}(x),  
    \label{charge-spin}
\end{split}
\end{equation}
where $\hat{Q}^{(\alpha)}_i/\hat{S}^{(\alpha)}_i$ is the charge/spin operator defined in the $SC_i/FM_i$ region, with eigenvalues as $q_i^{(\alpha)}/s_i^{(\alpha)}$, where $\alpha$ references the $\phi_{\alpha}$ boson mode.

We take the limit $\Delta_{\alpha},\mathcal{M}_{\alpha} \rightarrow \infty$, so that all the cosines in Eq. \ref{Eq:pairing} are pinned to their respective minima. These minima are characterized by integer-valued operators,
\begin{eqnarray}
(\varphi_1 + \varphi_2)\big|_{SC_i} &=& \pi \hat{N}^{\varphi_1+\varphi_2}_{i};~
\varphi_2\big|_{SC_i} = \frac{2\pi}{3}\hat{N}^{\varphi_2}_{i}  \nonumber\\
(\theta_1 + \theta_2)\big|_{FM_i} &=& \pi \hat{N}^{\theta_1+\theta_2}_{i};~
\theta_2\big|_{FM_i} = \frac{2\pi}{3}\hat{N}^{\theta_2}_{i} 
\label{Eq:BC}
\end{eqnarray}
and  one can also get $\varphi_{1}\big|_{SC_{i}}$ and $\theta_{1}\big|_{SC_{i}}$ from Eq.~\ref{Eq:BC}. $\hat{N}^{\mu}_{i}$ denotes the integer-valued minima to which the field $\mu$ is pinned in the $i$-th SC (for the $\varphi$ fields) or FM (for the $\theta$ fields) regions. They can take values $\hat{N}_{i}^{\varphi_1+\varphi_2/\theta_1+\theta_2} \in \lbrace 0,1 \rbrace$, $\hat{N}^{\varphi_1/\theta_1}_{i} \in \lbrace 0,\dots,5 \rbrace$ and $\hat{N}^{\varphi_2/\theta_2}_{i} \in \lbrace 0,1,2 \rbrace$, where the possible values are dictated by the number of physically distinct minima admitted by the respective cosines.

The operators $\hat{N}^\mu_i$ and the charge/spin operators defined in Eq. \ref{charge-spin} are related by \cite{iyer23,snizhko18} 
\begin{eqnarray}
\hat{Q}_{i}/\hat{S}_{i} &=&  \hat{N}^{\varphi_{1}+\varphi_{2}/\theta_{1}+\theta_{2}}_{i+1}-\hat{N}^{\varphi_{1}+\varphi_{2}/\theta_{1}+\theta_{2}}_{i}\nonumber\\
\hat{Q}^{(\alpha)}_i/\hat{S}^{(\alpha)}_i &=& \frac{\alpha}{3}\left(\hat{N}^{\theta_\alpha/\varphi_\alpha}_{i+1}- \hat{N}^{\theta_\alpha/\varphi_\alpha}_{i} \right)
\label{QS-N_reln}
\end{eqnarray}
where $\hat{Q}_{i}/\hat{S}_{i} = \sum_\alpha\hat{Q}^{(\alpha)}_{i}/\hat{S}^{(\alpha)}_{i}$ is the total charge/spin operator for the $SC_i/FM_i$ region. It is then clear that
\begin{eqnarray}
    \hat{N}^{\varphi_{1} + \varphi_{2}/\theta_{1} + \theta_{2}}_{j} &=& \mathrm{Mod}\left[\frac{1}{3}\hat{N}^{\varphi_{1}/\theta_{1}}_{j} + \frac{2}{3}\hat{N}^{\varphi_{2}/\theta_{2}}_{j},2\right]
\end{eqnarray}
that is, only those values of $\hat{N}^{\varphi_{\alpha}/\theta_{\alpha}}_{j}$ are allowed in the system such that $\hat{N}_{j}^{\varphi_1+\varphi_2/\theta_1+\theta_2} \in \lbrace 0,1 \rbrace$. 
From Eq. \ref{QS-N_reln}
it follows that only integer eigenvalues of ${\hat Q}_j$/${\hat S}_j$ are allowed in $SC_j/FM_j$. That is, despite the existence of a $\nu = 1/3$ bulk, the reconstructed edge always hosts only an integer charge/spin on $SC_j/FM_j$. Hence, the total charge/spin on the reconstructed edge is also an integer. For non-integer values of charge/spin on $SC_j/FM_j$, the system moves away from the simultaneous minima of the cosines in Eq. \ref{Eq:pairing}.

The charge $e^{i\pi\hat{Q_{i}}}$ (in $SC_{i}$) and spin $e^{i\pi\hat{S}_{i}}$ (in $FM_{i}$) parity operators obey global constraints set  by the number of flux quantum in the QH bulks
\begin{equation}
\begin{split}
&\prod_i e^{i\pi\hat{Q}_{i}} = e^{i\pi \left( n_{\uparrow}+n_{\downarrow} + \frac{1}{3}(m_{\uparrow}+m_{\downarrow})\right)} \\
& \prod_i e^{i\pi\hat{S}_{i}} =  e^{i\pi \left( n_{\uparrow}-n_{\downarrow} + \frac{1}{3}(m_{\uparrow} - m_{\downarrow})\right)} ,
    \label{Eq:Global_constraint}
\end{split}    
\end{equation} 
where $n_{\uparrow/\downarrow} (m_{\uparrow/\downarrow})$ denotes the number of flux quanta in the up/down-spin $\nu_{\uparrow/\downarrow}=1$($1/3$) bulk. We can now define $q_{totA}/s_{totA} = n_\uparrow \pm n_\downarrow$ as the total charge/spin in the $\nu = 1$ bulk, and  $q_{totB}/s_{totB} = (m_\uparrow\pm m_\downarrow)/3$ as the total charge/spin in the $\nu = 1/3$ bulk. The subscripts $A/B$ remind us that fluxes in the $\nu = 1 (1/3)$ bulk, by the bulk-edge correspondence, appear on the edge as $\psi_A (\psi_B)$ electrons, as mentioned earlier. The total charge/spin of the system, denoted by $q_{tot}/s_{tot}$, is the sum of the total charge/spin on $\nu =1 $ and $1/3$ bulks. By definition, $q_{totA}/s_{totA} \in \{0,1\}$ since it is the charge in the $\nu=1$ bulk, and we showed earlier that the total charge/spin on the edge has to be an integer, it follows that $q_{totB}/s_{totB} \in \{0,1\}$. 

The total charge and spin parity of the edges in the SC and FM regions characterize the ground state of the system, and the number of ways these can be distributed in the different SC and FM regions gives the degeneracy of the ground states. From Eq. \ref{Eq:Global_constraint}, $(q_{totA}, s_{totA})$ are constrained to be simultaneously even or odd, and so are $(q_{totB}, s_{totB})$, but the two sets are independent
of each other. An unreconstructed edge (with no $\nu = 1/3$ side strip) has $q_{totB} = s_{totB} = 0$, giving us $2$ distinct parity solutions to the constraints of Eq. \ref{Eq:Global_constraint}, consistent with Ref. \cite{Lindner2012_parafermions}. For a reconstructed system, the number of distinct solutions is doubled compared to an unreconstructed system, since $q_{totA}$ and $q_{totB}$ can be even or odd independently.
\iffalse
\sout{To count the degeneracy of the ground state manifold, we use a set of mutually commuting operators that count the number of spins due to $\psi_A$ and $\psi_B$ electrons in $FM_1$ and $FM_2$, and the total charge due to the two electrons. This is given by the set of operators: $\Big\{e^{\frac{3i\pi}{2}\hat{S}_1^{(2)}},~e^{ i\pi\left(\hat{S}_1^{(1)}-\hat{S}_1^{(2)}/2\right)},~e^{\frac{3i\pi}{2}\hat{S}_2^{(2)}},~e^{ i\pi\left(\hat{S}_2^{(1)}-\hat{S}_2^{(2)}/2\right)},$ $~e^{\frac{3i\pi}{2}\hat{Q}^{(1)}_\text{tot}},~e^{i\pi\left(\hat{Q}^{(1)}_\text{tot}-\hat{Q}^{(2)}_\text{tot}/2\right)} \Big\}$ denoting each state by  $\ket{ s_{1A},s_{1B};~s_{2A},s_{2B};~q_{totA},q_{totB}}$. Here} 
 \sout{are the eigenvalues of the above operators. We emphasize that $s_{jA/B}$ count the spins due to the $A/B-$type electron on the $j-$th FM region, and $q_{totA/B}$, the total charge due to the $A/B-$type electron \cite{supplemental}. From the allowed values of $s_j^{(\alpha)},q_{tot}$, we can show $s_{jA/B}, q_{totA/B} \in \{0, 1 \}.$}
\fi
  
The degeneracy of the system, for a fixed $q_{totA}$ and $q_{totB}$, can be counted by considering the possibilities of distributing the spin parity due to $A$ and $B$ type electrons on the different FM regions. Each state can then be denoted by $\ket{ s_{1A},s_{1B};~s_{2A},s_{2B};~q_{totA},q_{totB}}$ \cite{[{See Supplemental Material at }][{ for details on theory of reconstructed edge, counting the degeneracy of the ground states, and calculation of the Josephson junction.}]supp}
with
\begin{equation}
\begin{split}
 s_{jA} = 3s^{(2)}_j/2; ~ s_{jB} &= s_j^{(1)}-s_j^{(2)}/2 \\
\end{split}
\end{equation}
$s_{jA/B}$ (defined mod(2)) counts the spin parity due to the $A/B$ type of electrons on the $FM_{j}$ region. From the allowed values of $s_j^{(\alpha)}$, we can show $s_{jA/B} \in \{0, 1 \}.$ That is, the spin parity due to each electron can be 0 or 1 in each FM region. This, together with the constraint of Eq. \ref{Eq:Global_constraint} that the total charge and spin due to each electron have to be simultaneously even or odd, leads to $4$ degenerate states for a fixed $q_{totA}$ and $q_{totB}$ \cite{[{See Supplemental Material at }][{ for details on theory of reconstructed edge, counting the degeneracy of the ground states, and calculation of the Josephson junction.}]supp}. Note again that for an unreconstructed edge, the spin and charge due $\psi_B$ electron would be absent, giving back $2$ degenerate states for a fixed $q_{totA}.$

Commutation relations between the charge and spin operators \cite{[{See Supplemental Material at }][{ for details on theory of reconstructed edge, counting the degeneracy of the ground states, and calculation of the Josephson junction.}]supp} imply that the action of $e^{i\pi\hat{Q}_i}$ on the ground states transfers a spin due to $\psi_A$-type electron from $FM_i$ to $FM_{i+1}$. That is, $e^{i\pi\hat{Q}_1}\ket{ s_{1A},s_{1B};s_{2A},s_{2B};q_{totA},q_{totB}} = \ket{ s_{1A}-1,s_{1B};s_{2A}+1,s_{2B};q_{totA},q_{totB}}$. Similarly, the operator $e^{i3\pi\hat{Q}_i^{(1)}}$ transfers a spin due to $\psi_B-$type electron from $FM_i$ to $FM_{i+1}$: $e^{3i\pi\hat{Q}^{(1)}_1}\ket{ s_{1A},s_{1B};s_{2A},s_{2B};q_{totA},q_{totB}} = \ket{ s_{1A},s_{1B}-1;s_{2A},s_{2B}+1;q_{totA},q_{totB}}$, taking the system between states which cannot be accessed by the action of $e^{i\pi\hat{Q}_i}$. Acting both these operators twice gives back the same state. Hence, the ground state manifold containing $4$ states decouples under the action of the operators $e^{i\pi\hat{Q}_i}$ and $e^{3i\pi\hat{Q}^{(1)}_i}$ into $2$ disjoint submanifolds containing $2$ states each. With the combined action of $e^{i\pi\hat{Q}_i}$ and $e^{i3\pi\hat{Q}_i^{(1)}}$, the system can be rotated into any state in the ground state manifold.

\emph{Interface operators:--} 
The doubling of the ground state degeneracy due to the possibility of forming two types of electrons on the edge ($\psi_{A}$ and $\psi_{B}$) implies the existence of two Majorana zero modes at the SC-FM interfaces. The two Majorana operators are denoted by $\chi_{2i-1,\sigma}$ and $\alpha_{2i-1,\sigma}$, at the interface of $SC_{i}$ and $FM_{i}$, and $\chi_{2i}$ and $\alpha_{2i}$ at the interface of $FM_{i}$ and $SC_{i+1}$, and are given by
\begin{eqnarray}
    \chi_{2i-1,\sigma} &=& e^{i\sigma (\theta^{(i)}_{1} + \theta^{(i)}_{2})} \hat{T}^{C}_{1}\hat{T}^{C}_{2}\Pi_{j=1}^{i}e^{i\pi\hat{S}_{j-1}} \nonumber\\
     \chi_{2i,\sigma} &=& e^{i\sigma (\theta^{(i)}_{1} + \theta^{(i)}_{2})} \hat{T}^{C}_{1}\hat{T}^{C}_{2}\Pi_{j=1}^{i}e^{i\pi\hat{S}_{j}} \nonumber\\
     \alpha_{2i-1,\sigma} &=& e^{3i\sigma \theta^{(i)}_{1}} (\hat{T}^{C}_{1})^{3}\Pi_{j=1}^{i}e^{3i\pi\hat{S}^{(1)}_{j-1}} \nonumber\\
    \alpha_{2i,\sigma} &=& e^{3i\sigma \theta^{(i)}_{1}} (\hat{T}^{C}_{1})^{3}\Pi_{j=1}^{i}e^{3i\pi\hat{S}^{(1)}_{j}} 
\end{eqnarray}
where $\hat{T}^{C}_{1}$ and $\hat{T}^{C}_{2}$ increase the total charge 
on the $1/3$ and the $2/3$ bosonic modes by $e/3$ and $2e/3$, respectively. The operator $e^{p i\sigma\theta^{(i)}_{\alpha}}$ adds a spin $p \sigma \nu_{\alpha}$ on $FM_i$. Since $\psi_A$ and $\psi_B$ cannot transform into each other, reflected in the ground-state manifold being decoupled into two disjoint parts, these two Majoranas are decoupled although they exist at the same physical location. The two decoupled Majoranas can form two distinct resonant levels with their counterparts on opposite interfaces.

\emph{Josephson Junction:--} Now we take the limit $\mathcal{M}_1\rightarrow0$, realizing a Josephson junction between the two SC as shown in Fig. \ref{fig:set_up}. The system is described by the Hamiltonian, $H =  H_0 + H_{SC} + H_{FM}$ as given in Eq.~\ref{Eq:pairing}.
The pairing Hamiltonian in the presence of a finite SC phase bias is given by 
\begin{equation}
\begin{split}
    H_{SC} = -\sum_{\gamma=A,B,C} \Delta_\gamma \left(\int_{ SC_1} dx~ \psi_\gamma \psi_\gamma +\right.\\
    \left.\int_{SC_2 }dx~  e^{i\phi_{SC}} \psi_\gamma \psi_\gamma + hc \right)
    \label{Eq:JJ_pairing} 
\end{split}    
\end{equation}
where the SC phase difference, $\phi_{SC}=\phi^{SC}_{2}-\phi^{SC}_{1}$ is plugged into $SC_2$ using gauge invariance. The above terms translate into cosine operators of the bosonic fields \cite{[{See Supplemental Material at }][{ for details on theory of reconstructed edge, counting the degeneracy of the ground states, and calculation of the Josephson junction.}]supp} which in the limit, $\Delta_\gamma \rightarrow \infty$, are pinned to their minima, imposing boundary conditions on the bosonic fields in $H_0$
\begin{eqnarray}
  (\varphi_1 + \varphi_2)(L_1) &=& 0;~ \varphi_{1}(L_1) = 0;~ \varphi_{2}(L_1) = 0 \nonumber\\
  (\varphi_1 + \varphi_2)(L_2) &=& \hat{F};~ \varphi_{1}(L_2) = \hat{F}_1;~ \varphi_{2}(L_2) = \hat{F}_2
    \label{boundary_conditions}
\end{eqnarray}
\begin{gather}
   \hat{F} = \mathrm{mod}\left[\pi\left(\frac{\hat{N}^{\varphi_1}_{2}}{3} + \frac{2\hat{N}^{\varphi_2}_{2}}{3} - \frac{\phi_{SC}}{2\pi}\right) + \pi,2\pi\right] - \pi \nonumber\\
    \hat{F}_{2} = \frac{2}{3}\left( \mathrm{mod}\left[ \pi \left(\hat{N}^{\varphi_2}_{2}-\frac{\phi_{SC}}{2\pi}\right)+\pi,2\pi\right]-\pi\right)\nonumber\\
    \hat{F}_1 = \hat{F}-\hat{F}_2
    \label{compactness}
\end{gather}
where the mods are imposed to ensure single-valuedness of the electron operators and spins $s_{1}^{(1)}$ and $s_{1}^{(2)}$ to be consistent with the state of the system \cite{Clarke2013_parafermions,iyer23}. The bosonic fields $\varphi_j$ and $\theta_j$ are then mode-expanded to satisfy the boundary conditions and used to diagonalize the Hamiltonian. This gives the low energy part of the Josephson spectrum to be \cite{supp} 
\begin{eqnarray}
    H &=& \sum_{\alpha=1}^{2}\frac{\hbar v_{\alpha}  \left(\hat{\varphi}_{\alpha}(L_{2})\right)^{2}}{2\nu_{\alpha}\pi (L_2-L_1)}. 
    \label{Eq:hamiltonian_2}
\end{eqnarray} 
\begin{figure}
\centering
    \includegraphics[scale=0.32]{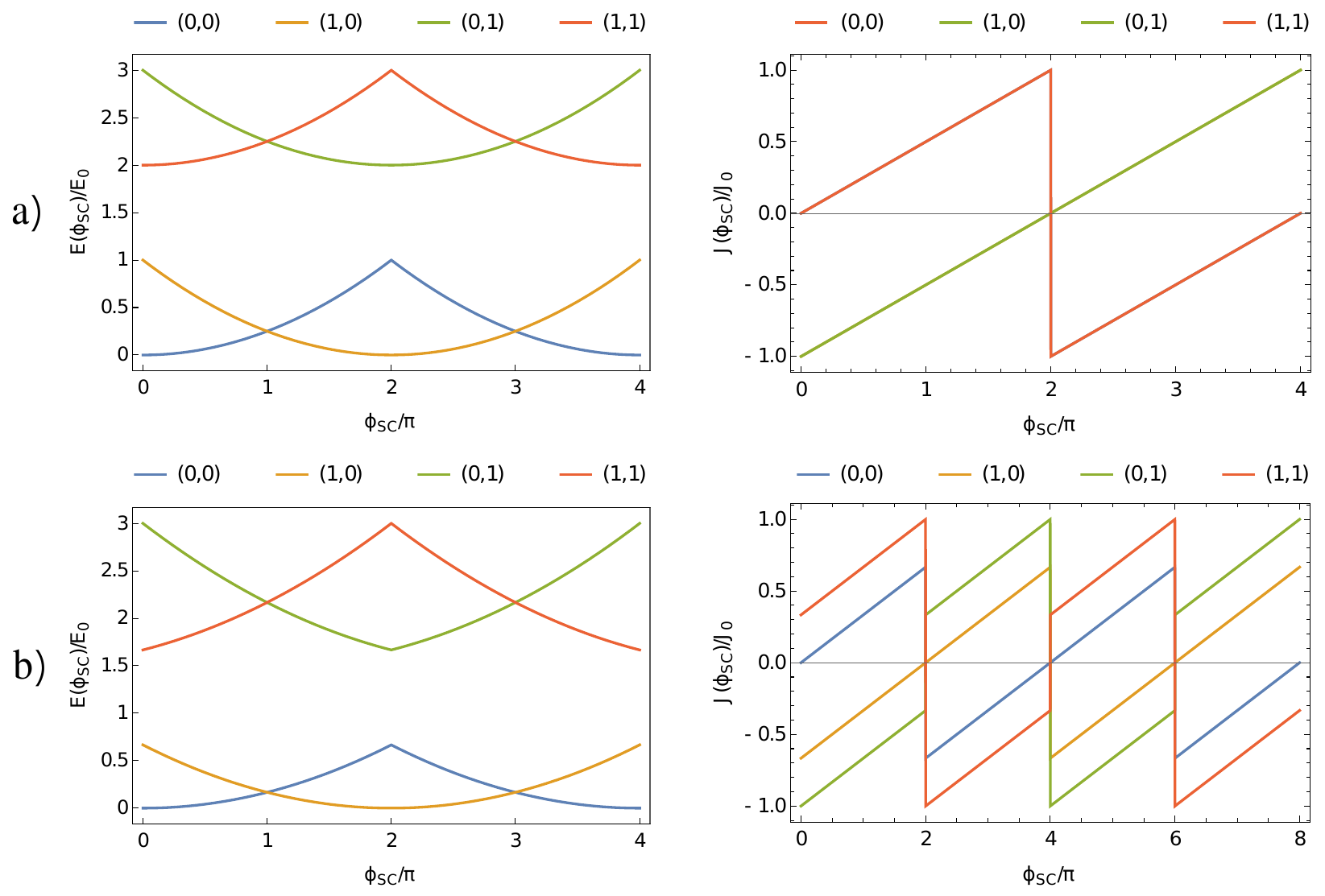}
    \caption{
    The energy spectrum ($E(\phi_{SC})$) and the corresponding Josephson current ($J(\phi_{SC})$) is plotted as a function of SC phase bias for different values of $(s_{1A},s_{1B})$. a) shows $E(\phi_{SC})$ and $J(\phi_{SC})$ for equal propagation velocities of the bosonic modes, $v_{1} = v_{2} = 1 $. b) shows $E(\phi_{SC})$ and $J(\phi_{SC})$ when the bosonic modes have different propagation velocities, $v_{1} = 0.5 v_2 = 1$. The energy eigenvalues  ($E(\phi_{SC})$) and Josephson current ($J(\phi_{SC})$) are normalized with respect to the maximum energy ($E_{0}$) and Josephson current ($J_{0}$) of the $\nu=1$ unreconstructed edge.  }
    \label{fig:spectrum_JJC}
\end{figure} 
The energy eigenvalue depends only on the spin parity of different electrons contained in the Josephson junction and is given by $H\ket{s_{1A},s_{1B};s_{2A},s_{2B};q_{totA},q_{totB}} = E\left(s_{1A},s_{1B}\right)\ket{s_{1A},s_{1B};s_{2A},s_{2B};q_{totA},q_{totB}}$. The Josephson current $J(\phi_{SC})$, is given by $d\langle H \rangle/d \phi_{SC}$, which for $v_1 = v_2$, is sensitive only to the full spin $s_{1A} + s_{1B}$ parity of the junction. Hence, in Fig. \ref{fig:spectrum_JJC}b there exist overlapping curves corresponding to different spin parities of the different electron types but the same total parities. This feature of the $J_{\phi_{SC}}$ is broken for $v_1 \neq v_2$, where each state distinguished by the spin parity of different electron types has a distinct signature. We show this in Fig. \ref{fig:spectrum_JJC}d for $v_{2}/v_{1} = 0.5$ where the $J(\phi_{SC})$ in all the different states in the ground-state manifold displays distinct signatures.

\emph{Discussion and conclusions:--} In this letter, we have studied the effect of edge reconstruction on Majorana zero modes formed on the edges of a $\nu = 1$ quantum Hall system proximitized by superconductors and ferromagnets. Edge reconstruction-induced deposition of an additional $\nu =1/3$ quantum Hall side strip leads to the possibility of multiple types of electrons existing on the edges. At each SC-FM interface, we find two decoupled Majorana modes that are concerned with the occupancy of two different type of electrons: one related to the $\nu = 1$ bulk excitations, and the other to the $\nu = 1/3$ bulk excitations. It is important to emphasize that the $\nu = 1/3$ bulk does not give rise to a parafermion, but a Majorana as the fractional occupation of the parafermion is energetically unfavorable.  

We propose measuring the Josephson current to probe the existence of two decoupled Majoranas at each interface. When the velocities of the $1/3$ and $2/3$ bosonic modes are equal, the fractional Josephson current does not distinguish between the states with the same junction spin parity. When the bosonic mode velocities are different, all the states in the ground-state manifold display distinct signatures in the fractional Josephson current. In all the above scenarios, however, the periodicity of the Josephson current remains $4\pi$ with the SC phase bias.

\begin{acknowledgments}
A.R. would like to thank Efrat Shimshoni and Ady Stern for useful discussions. A.R. also acknowledges the University Grants Commission, India, for support in the form of a fellowship. S.D. acknowledges the warm hospitality from ICTS during the final stages of writing the draft. 
K.I. acknowledges support from CEFIPRA through the Raman-Charpak Fellowship, and thanks ICTS for hospitality. 
\end{acknowledgments}

\vspace{+0.2cm}
\emph{Author contribution:-}
The first two authors, K.I.  and A.R. have contributed equally to this work. \\

\clearpage
\widetext{
\begin{center}
\textbf{\large Supplemental material for ``Majorana zero modes in quantum Hall edges survive edge reconstruction"}
\end{center}
\numberwithin{equation}{section}

\section{Chiral Luttinger Liquid Theory of the Reconstructed $\nu = 1$ quantum Hall edge}

In this section, we describe the chiral Luttinger liquid theory of a reconstructed quantum Hall edge belonging to the $\nu=1$ bulk. Our starting point is a general edge theory of Abelian quantum Hall systems of filling fraction $\nu$ with an arbitrary edge structure, following Ref. \cite{shtanko14}. The edge action of a quantum Hall system carrying $N$ bosonic edge modes on the edge is given by
\begin{equation}
    S = \frac{1}{4\pi}\int dx dt \sum_{i=1}^{N}\left( -\chi_{i}\partial_{x}\phi_{i}\partial_{t}\phi_{i} -v_{i}\left(\partial_{x}\phi_{i}\right)^{2}\right)
    \label{edge}
\end{equation}
where $\phi_{i}$ is the bosonic field corresponding to the $i^{th}$ edge mode, $v_{i}$ denote the propagation velocity of the  $i^{th}$ bosonic mode and $\chi_i = \pm 1$ represents the chirality of the $i^{th}$ bosonic mode. The conductance of the $i^{th}$ mode mode is given by $q_{i}^{2}$, which must satisfy a constraint determined by the bulk filling factor $\nu$
\begin{equation}
    \sum_{i=1}^{N} \chi_{i}q_{i}^{2} = \nu.
\end{equation}
The total current and charge density on the edge is given by $\rho(x,t) = \frac{1}{2\pi}\sum_{i=1}^{N} q_{i}\partial_{x}\phi_{i}(x,t)$ and $J(x,t) = -\frac{1}{2\pi}\sum_{i=1}^{N} q_{i}\partial_{t}\phi_{i}(x,t)$ respectively, such that the bosonic fields follow the continuity equation 
$\partial_{t} \rho(x,t) + \partial_{x}J(x,t) = 0$.

Bosonic field $\phi_{i}$ quantized over an edge of length $L$ can be expressed as
\begin{eqnarray}
    \phi(x,t) &=& \hat{\phi}_{i}^{0} + \frac{2\pi}{L}q_{i}\hat{N}_{i} X_{i}  + i\sum_{n=1}^{\infty} \sqrt{\frac{2\pi}{L k}}\left(\hat{a}_{ik}e^{-ikX_{i}} - \hat{a}^{\dagger}_{ik}e^{ikX_{i}}\right)
\end{eqnarray}
where $X_{i} = -\chi_{i}x + v_{i}t$, $k = 2\pi n/L$, $n \in \mathcal{N}$, $\hat{a}_{ik}$ is the bosonic annihilation operator and $\hat{\phi}_{i}^{0}$ and $\hat{N}_{i}$ are the zero mode phase and number operator, with commutation relation $\left[\hat{a}_{i,k},\hat{a}_{j,k'}^{\dagger}\right] = \delta_{ij}\delta_{kk'}$ and $\left[\hat{\phi}_{i}^{0},\hat{N}_{j}\right] = -i \delta_{ij}$, such that the bosonic fields follows the commutation relation given by $\left[\phi_{i}(x,t),\phi_{j}(x',t')\right] = -i\pi \mathrm{sgn}(X_{i}-X'_{j})\delta_{ij}$.

A general 1-D electronic field can be represented in terms of bosonic fields as
\begin{eqnarray}
    \psi_{\alpha}(x,t) &=& \left(\frac{L}{2\pi}\right)^{-\sum_{i}e_{\alpha,i}^{2}/2} e^{i\sum_{i}e_{\alpha,i} \phi_{i}(x,t)}.
\end{eqnarray}
which satisfy the constraints 
\begin{eqnarray}
    \lbrace \psi_{\alpha}(x,t),\psi_{\alpha}(x',t) \rbrace &=& 0, \nonumber\\
    \psi_{\alpha}(x,t)\psi_{\beta}(x',t) \pm \psi_{\beta}(x',t),\psi_{\alpha}(x,t) &=& 0, \nonumber\\
    \left[\rho(x,t),\psi_{\alpha}(x',t)\right] &=& \delta(x-x')\psi_{\alpha}(x,t). \nonumber\\
\end{eqnarray}
In terms of the parameters $e_{\alpha,i}$, the constraints are expressed as \begin{eqnarray}
\textbf{e}_{\alpha}.\textbf{e}_{\alpha} &\in& 2\mathrm{Z}+1, \nonumber\\
    \textbf{e}_{\alpha}.\textbf{e}_{\beta} &\in& \mathrm{Z}, \nonumber\\
    \textbf{q}.\textbf{e}_{\alpha} &=& -1,
    \label{Eq:fermion_field_parameter}
\end{eqnarray}
where $\textbf{e}_{\alpha} = (e_{\alpha,1},....,e_{\alpha,N})$, $\textbf{q} = (q_{1},.....,q_{N})$ and the operation $\textbf{A}.\textbf{B}$ is defined as $\textbf{A}.\textbf{B} = \sum_{i=1}^{N}\chi_{i}A_{i}B_{i}$. The \textbf{K} matrix for the edge is given by $\textbf{K}_{\alpha \beta} = \textbf{e}_{\alpha}.\textbf{e}_{\beta}$.

Now, we use the above formalism to find the electron operators of an edge reconstructed $\nu =1$ quantum Hall system. Such a system consists of three bosonic modes on its edge: two downstream charged modes with conductance 1/3 and 2/3, and a third upstream neutral bosonic mode \cite{khanna_fractional_2021}. The edge is then described by the charge vector $\textbf{q} = \left(\sqrt{\frac{1}{3}},\sqrt{\frac{2}{3}},0\right)$, and the chirality vector $\boldsymbol{\chi} = (1,1,-1)$. Together with the action of Eq. \ref{edge}, these numbers fully describe the edge theory of the reconstructed $\nu = 1$ system. Solving for $\textbf{e}_{\alpha}$ with these inputs of $\textbf{q}$ and $\boldsymbol{\chi}$, we obtain the following electron operators
\begin{eqnarray}
  \psi_{A, m}(x,t) &\sim& e^{-i\sqrt{\frac{1}{3}} \phi_{1} -i\sqrt{\frac{2}{3}}\phi_{2}\pm i \sqrt{2m}\phi_{3}}, \nonumber\\
  \psi_{B,m}(x,t) &\sim& e^{-i\sqrt{3} \phi_{1} \pm i \sqrt{2-2m}\phi_{3}}, \nonumber\\
  \psi_{C,m}(x,t) &\sim& e^{-i\sqrt{\frac{3}{2}} \phi_{2} \pm i\sqrt{\frac{1}{2}-2m}\phi_{3}}.
  \label{Eq:All_electron_operators}
 \end{eqnarray}
where $m$ denotes an positive integer. Here, $\psi_{A,m}$ denotes electron fields with composed of excitations on all the three bosonic fields, while $\psi_{B/C,m}$ comprises excitations only on the $\phi_{1/2}$ fields and the neutral mode. In all these classes of electron operators, the ones with the lowest scaling dimension are given by 
\begin{eqnarray}
  \psi_{A}(x,t) &\sim& e^{-i\sqrt{\frac{1}{3}} \phi_{1}-i\sqrt{\frac{2}{3}}\phi_{2}} \nonumber\\
  \psi_{B}(x,t) &\sim& e^{-i\sqrt{3} \phi_{1}} \nonumber\\
  \psi_{C}(x,t) &\sim& e^{-i\sqrt{\frac{3}{2}} \phi_{2} \pm i\sqrt{\frac{1}{2}}\phi_{3}}.
  \label{Eq:All_electron_operators}
 \end{eqnarray}
In the main text, we use a notation where the conductances of the bosonic edges appear as coefficients in the action (but not in the definition of charge/current operators.) Then, the commutators of the bosonic fields also gain the conductance as a coefficient to the sign function, becoming: $\left[\phi_{i}(x,t),\phi_{j}(x',t')\right] = -i\pi \nu_{i}\delta_{ij}\mathrm{sgn}(X_{i}-X'_{j})$ as given in the main text. In this notation, the electron operators are given by 
\begin{eqnarray}
  \psi_{A}(x,t) &\propto& e^{i\phi_{1}+i\phi_{2}} \nonumber\\
  \psi_{B}(x,t) &\propto& e^{i 3\phi_{1}} \nonumber\\
  \psi_{C}(x,t) &\propto& e^{i\frac{3}{2} \phi_{2} \pm i\frac{1}{2}\phi_{3}}. \label{Eq:relevant_electron_operators}
\end{eqnarray}
which is the form of these operators used in the main text.

\section{Commutation relations}
Charge on the $\alpha-$th bosonic mode of $SC_{1}$ and $SC_{2}$ is given by
\begin{eqnarray}    \hat{Q}^{(\alpha)}_{1} &=& \frac{1}{\pi}\left( \theta_{\alpha}(x_{1}) - \theta_{\alpha}(0) + \theta_{\alpha}(L)-\theta_{\alpha}(x_{4})\right) \nonumber\\
\hat{Q}^{(\alpha)}_{2} &=& \frac{1}{\pi} \left( \theta_{\alpha}(x_{3}) - \theta_{\alpha}(x_{2}) \right), 
\end{eqnarray}
with $\hat{Q}_j = \sum_{\alpha=1,2} \hat{Q}^{(\alpha)}_{j}$ giving the total charge encompassed in the $SC_j$ region. Similarly, spin content of $FM_{1}$ and $FM_{2}$ is given by
\begin{eqnarray}
    \hat{S}^{(\alpha)}_{1} &=& \frac{1}{\pi}\left( \phi_{\alpha}(x_{2}) - \phi_{\alpha}(x_{1})
 \right) \nonumber\\
 \hat{S}^{(\alpha)}_{2} &=& \frac{1}{\pi} \left( \phi_{\alpha}(x_{4}) - \phi_{\alpha}(x_{3})
 \right).
\end{eqnarray}
with $\hat{S}_j = \sum_{\alpha=1,2} \hat{S}^{(\alpha)}_{j}$ being the full spin within $FM_j$. 

The commutation relation of the bosonic fields, $\left[\phi_\alpha(x),\theta_\beta(y)\right] = i\pi \nu_\alpha \delta_{\alpha\beta} \Theta (x-y)$, where $\nu_\alpha = \alpha/3, \alpha \in \{1, 2\}$ leads to the following relations for the charge and spin operators
\begin{equation}
\begin{split}
 \left[ 
\hat{Q}^{(\alpha)}_{1},\hat{Q}^{(\beta)}_{2} \right] &= \left[ 
\hat{S}^{(\alpha)}_{1},\hat{S}^{(\beta)}_{2} \right] = 0    \\
\left[ 
\hat{Q}^{(\alpha)}_{1},\hat{S}^{(\beta)}_{1} \right] &= \left[ 
\hat{Q}^{(\alpha)}_{2},\hat{S}^{(\beta)}_{2} \right] = -\frac{i\nu_\alpha}{\pi}\delta_{\alpha\beta} \\
\left[ 
\hat{Q}^{(\alpha)}_{2},\hat{S}^{(\beta)}_{1} \right] &= \left[ 
\hat{Q}^{(\alpha)}_{1},\hat{S}^{(\beta)}_{2} \right] = \frac{i\nu_\alpha}{\pi}\delta_{\alpha\beta}
\end{split}    
\end{equation}
From the above, the commutation relation between the charge and spin parity operators is given by
\begin{eqnarray}
\left[e^{i\pi\hat{Q}^{(\alpha)}_{1}},e^{i\pi\hat{Q}^{(\beta)}_{2}}\right] &=& \left[e^{i\pi\hat{S}^{(\alpha)}_{1}},e^{i\pi\hat{S}^{(\beta)}_{2}}\right] = 0 \nonumber\\
\left[e^{i\pi\hat{Q}^{(\alpha)}_{i}},e^{i\pi\hat{S}^{(\beta)}_{tot}}\right] &=& \left[e^{i\pi\hat{S}^{(\alpha)}_{i}},e^{i\pi\hat{Q}^{(\beta)}_{tot}}\right] = 0 \nonumber\\
e^{i\pi\hat{Q}^{(\alpha)}_{i}}e^{i\pi\hat{S}^{(\alpha)}_{j}} &=& e^{i\pi\nu_\alpha\left( \delta_{i,j}-\delta_{i+1,j} \right)}e^{i\pi\hat{S}^{(\alpha)}_{j}}e^{i\pi\hat{Q}^{(\alpha)}_{i}}
\label{parity_commutators}
\end{eqnarray}
where $\hat{S}^{(\beta)}_{tot} = \sum_j \hat{S}^{(\beta)}_j$. 

\section{Bases for the ground state manifold}

We consider the system in the limit of $\Delta_\gamma, \mathcal{M}_\gamma$ where all the cosines in Eq. of the main text are pinned to their minima. To label the states in the ground state manifold, it seems natural to diagonalize the Hamiltonian in the basis of the following mutually commuting operators which count the number of fundamental excitations on each bosonic field
\begin{equation}
\left\{e^{i\pi\hat{S}_1^{(1)}},~e^{i\pi\hat{S}_1^{(2)}},~e^{i\pi\hat{S}_2^{(1)}},~e^{i\pi\hat{S}_2^{(2)}},~e^{i\pi\hat{Q}^{(1)}_\text{tot}},~e^{i\pi\hat{Q}^{(2)}_\text{tot}} \right\}
\end{equation}
These operators are characterized by eigenvalues
\begin{equation}
    \left\{ e^{i\pi s_1^{(1)}},~e^{i\pi s_1^{(2)}},~e^{i\pi s_2^{(1)}},~e^{i\pi s_2^{(2)}},~e^{i\pi q^{(1)}_{\text{tot}}}, e^{i\pi q^{(2)}_{\text{tot}}} \right\}
\end{equation}
which allows us to label eigenstates in the ground state manifold as
\begin{equation}
\ket{s_{1}^{(1)},~s_{1}^{(2)};~s_{2}^{(1)},~s_{2}^{(2)};~q^{(1)}_\text{tot},~q^{(2)}_\text{tot}}
\end{equation}
where $s^{(1)}_{j} \in \{0,1/3,\dots,5/3\}$, $s^{(2)}_{j} \in \{0,2/3,4/3\}$. However, to remain on the simultaneous minima of all the pinned cosines, the total spin on each FM region, $s_j^{(1)}+s_j^{(2)} = s_j$ must be an integer. This implies that $s_{tot} = s^{(1)}_{tot}+s^{(2)}_{tot} $ is an integer, and hence, so is $q_{tot} = q^{(1)}_{tot}+q^{(2)}_{tot}$. So, only those combinations of $\{s_{1}^{(1)},~s_{1}^{(2)},~s_{2}^{(1)},~s_{2}^{(2)},~q^{(1)}_\text{tot},~q^{(2)}_\text{tot}\}$ that add up to give integer $\{s_1, s_2, q_{\text{tot}}\}$ are allowed within the low energy approximation. The allowed values of the spins and charges are hence: $(s_j^{(1)},s_j^{(2)}), ~(q_\text{tot}^{(1)},q_\text{tot}^{(2)}) \in \{ (0,0),~(1/3,2/3),~(1,0),~(4/3,2/3) \}$, such that: $s_1, s_2, q_\text{tot} \in \{0, ~1 \}$

The operator $e^{i\pi\hat{Q}_j^{(1)}}$ transfers a spin 1/3 on the one-third bosonic mode from $FM_j$ region to $FM_{j+1}$ and $e^{i\pi\hat{Q}_j^{(2)}}$ transfers a spin 2/3 on the two-third bosonic mode from $FM_j$ region to $FM_{j+1}$. However, acting these operators individually takes the system out of the ground state manifold. There exist two combinations of these operators that when acted upon a state in the ground state manifold, takes it to yet another state in the manifold. These are: $e^{i\pi\hat{Q}_j^{(1)}}e^{i\pi\hat{Q}_j^{(2)}} = e^{i\pi\hat{Q}_j}$, and  $e^{3i\pi\hat{Q}_j^{(1)}}$. The action of these operators on an arbitrary state is as follows
\begin{equation}
\begin{split}
e^{i\pi\hat{Q}_1}\ket{s_{1}^{(1)},~s_{1}^{(2)};~s_{2}^{(1)},~s_{2}^{(2)};~q^{(1)}_\text{tot},~q^{(2)}_\text{tot}} &= \ket{s_{1}^{(1)}-1/3,~s_{1}^{(2)}-2/3;~s_{2}^{(1)}+1/3,~s_{2}^{(2)}+2/3;~q^{(1)}_\text{tot},~q^{(2)}_\text{tot}}  \\
e^{3i\pi\hat{Q}_1^{(1)}}\ket{s_{1}^{(1)},~s_{1}^{(2)};~s_{2}^{(1)},~s_{2}^{(2)};~q^{(1)}_\text{tot},~q^{(2)}_\text{tot}} &= \ket{s_{1}^{(1)}-1,~s_{1}^{(2)};~s_{2}^{(1)}+1,~s_{2}^{(2)};~q^{(1)}_\text{tot},~q^{(2)}_\text{tot}} 
\end{split}   
\end{equation}
where $e^{i\pi\hat{Q}_j}$ transfers an electronic spin of the $\psi_A-$type (composed of one excitation each on the 1/3 and 2/3 bosonic modes) and $e^{3i\pi\hat{Q}_j^{(1)}}$ transfers an electronic spin of the $\psi_B-$type (composed of three excitations on the 1/3 bosonic mode) from $FM_j$ to $FM_{j+1}$. The spin operators $e^{i\pi\hat{S}_1^{(1)}}$ and $e^{i\pi\hat{S}_1^{(2)}}$  simply read reads out the spin on the respective bosonic modes
\begin{equation}
\begin{split}
e^{i\pi\hat{S}_j^{(1)}}\ket{s_{1}^{(1)},~s_{1}^{(2)};~s_{2}^{(1)},~s_{2}^{(2)};~q_\text{tot}} &= e^{i\pi s_j^{(1)}}\ket{s_{1}^{(1)},~s_{1}^{(2)};~s_{2}^{(1)},~s_{2}^{(2)};~q_\text{tot}}  \\
e^{i\pi\hat{S}_j^{(2)}}\ket{s_{1}^{(1)},~s_{1}^{(2)};~s_{2}^{(1)},~s_{2}^{(2)};~q_\text{tot}} &= e^{i\pi s_j^{(2)}}\ket{s_{1}^{(1)},~s_{1}^{(2)};~s_{2}^{(1)},~s_{2}^{(2)};~q_\text{tot}}  \\
\end{split}   
\end{equation}

 \begin{center}
\begin{table}[h]
  \resizebox{0.60\textwidth}{!}{  
\begin{tabular}{||c| c| c| c| c| c| c||} 
 \hline
\multicolumn{1}{||c|}{$\nu=1$} & \multicolumn{2}{c|}{Sharp edge } & \multicolumn{4}{c||}{Reconstructed edge }\\ [0.5ex] 
 \hline\hline
 \multirow{2}{*}{Top. Sec. }  & \multicolumn{2}{c|}{$q_{totA}$ } & \multicolumn{4}{c||}{($q_{totA}$, $q_{totB}$) }\\ [0.5ex] \cline{2-7}
 %\hline
 
&0&1&(0, 0)
&(1, 1)&(1, 0)&(0, 1) \\ 
 \hline                                          
\multirow{4}{*}{\thead{} }   & \multicolumn{1}{c|}{\multirow{2}{*}{\textcolor{red}{$|0,0,0\rangle$}}} &  \multicolumn{1}{c|}{\multirow{2}{*}{}} & \textcolor{brown}{$|0,0,0,0,0,0\rangle$} & \textcolor{purple}{$|1,1;0,0;1,1\rangle$} & &\\  
                             &   & & \textcolor{brown}{$|1,0;1,0;0,0\rangle$} & \textcolor{purple}{$|0,1;1,0;1,1\rangle$} & & \\ \cline{4-7} 
                             & \multirow{2}{*}{\textcolor{red}{$|1,1,0\rangle$}}                       &  & \textcolor{olive}{$|1,1;1,1;0,0\rangle$} & \textcolor{magenta}{$|0,0;1,1;1,1\rangle$} &  & \\ 
                             &    &  & \textcolor{olive}{$|0,1;0,1;0,0\rangle$} & \textcolor{magenta}{$|1,0;0,1;1,1\rangle$} & & \\ \hline
\multirow{4}{*}{\thead{}}   & \multirow{2}{*}{}                       & \multirow{2}{*}{\textcolor{ForestGreen}{$|1,0,1\rangle$}}  &  & & \textcolor{blue}{$|1,0;0,0;1,0\rangle$} & \textcolor{violet}{$|0,1;0,0;0,1\rangle$}\\ 
                             &    &  &  & & \textcolor{blue}{$|0,0;1,0;1,0\rangle$} & \textcolor{violet}{$|1,1;1,0;0,1\rangle$}\\ \cline{4-7} 
                             & \multirow{2}{*}{}                       & \multirow{2}{*}{\textcolor{ForestGreen}{$|0,1,1\rangle$}} & & & \textcolor{MidnightBlue}{$|1,1;0,1;1,0\rangle$} & \textcolor{RawSienna}{$|0,0;0,1;0,1\rangle$} \\ 
                             &  &  & & & \textcolor{MidnightBlue}{$|0,1;1,1;1,0\rangle$} & \textcolor{RawSienna}{$|1,0;1,1;0,1\rangle$} \\ \hline
\end{tabular}
}
\caption{ The table exhaustively enumerates the ground state manifold of an edge-reconstructed $\nu =1$ system and contrasts it with that of a sharp or unreconstructed system. The fixed values of total charge parity, called the topological sector (Top. Sec.), for the unreconstructed edge is given by $q_{totA}$ and those for the reconstructed edge are given by ($q_{totA}$, $q_{totB}$). The states in the ground-state manifold are labelled as $\ket{s_{1A},s_{2A},q_{totA}}$ for the unreconstructed edge and as $\ket{s_{1A},s_{1B};s_{2A},s_{2B};q_{totA},q_{totB}}$ for the reconstructed edge. In a fixed topological sector, the unreconstructed system hosts $2$ states, while the edge reconstructed system hosts $4$ states. The eigenstates linked by the action of the operator $e^{i\pi \hat{Q}_{i}}$ are shown in the same colors.
}
\label{tab:Table1}
\end{table}
\end{center}

\noindent
Now we note an ambiguity in the current choice of basis. Assume we are in a state with one $\psi_A$ and one $\psi_B$ type spin on each FM. This state in the current notation is expressed as
\begin{equation}
    \ket{4,1;4,1;q^{(1)}_\text{tot},q^{(2)}_\text{tot}}
\end{equation}
Here, the numbers indicate the count of fundamental excitations in the ladder of each bosonic mode. However, given our knowledge of the structure of both electron operators, we can deduce that of the $4$ excitations on the $1/3$ ladder, $3$ come from $\psi_B$, and $1$ comes from $\psi_A$. The fact that $\psi_A$ has support on both bosonic modes ends up mixing the number of electrons with the numbers of excitations on the bosonic ladders, leading to an ambiguity.

We would like a basis where the numbers of each type of electron, $\psi_A$ and $\psi_B$, is transparent. For this, we note that $\psi_A$ electrons is always the same as the number of excitations on the $2/3$ ladder. One can leverage this information and transform it to a new basis, defined by a different set of mutually commuting set of operators:
\iffalse
\begin{equation}
\begin{split}
\left[\text{exp}\left(\frac{3i\pi}{2}\hat{S}_1^{(2)}\right), ~\text{exp}~ i\pi\left(\hat{S}_1^{(1)}-\frac{\hat{S}_1^{(2)}}{2
}\right),~\text{exp}\left(\frac{3i\pi}{2}\hat{S}_2^{(2)}\right),~\text{exp}~ i\pi\left(\hat{S}_2^{(1)}-\frac{\hat{S}_2^{(2)}}{2}\right), ~\text{exp}\left(\frac{3i\pi}{2}\hat{Q}^{(1)}_\text{tot}\right),~\text{exp}~ i\pi\left(\hat{Q}_{\text{tot}}^{(1)}-\frac{\hat{Q}_{\text{tot}}^{(2)}}{2}\right)\right] \nonumber\\
\equiv\left[e^{i\pi \hat{S}_{1A}},~e^{i\pi \hat{S}_{1B}},~e^{i\pi \hat{S}_{2A}},~e^{i\pi \hat{S}_{2B}},~e^{i\pi \hat{Q}_{totA}},~e^{i\pi \hat{Q}_{totB}}\right]
\end{split}
\end{equation}
\fi
\begin{equation}
\left\{\text{exp}\left(\frac{3i\pi}{2}\hat{S}_1^{(2)}\right),\text{exp}~ i\pi\left(\hat{S}_1^{(1)}-\frac{\hat{S}_1^{(2)}}{2
}\right),\text{exp}\left(\frac{3i\pi}{2}\hat{S}_2^{(2)}\right),\text{exp}~ i\pi\left(\hat{S}_2^{(1)}-\frac{\hat{S}_2^{(2)}}{2}\right),\text{exp}\left(\frac{3i\pi}{2}\hat{Q}^{(1)}_\text{tot}\right),~\exp{i\pi\left(\hat{Q}^{(1)}_\text{tot}-\frac{\hat{Q}^{(2)}_\text{tot}}{2}\right)} \right\}\nonumber
\end{equation}
\begin{equation}
    \equiv\left\{e^{i\pi \hat{S}_{1A}},~e^{i\pi \hat{S}_{1B}},~e^{i\pi \hat{S}_{2A}},~e^{i\pi \hat{S}_{2B}}~e^{i\pi \hat{Q}_{totA}},~e^{i\pi \hat{Q}_{totB}} \right\}
\end{equation}
\iffalse
\begin{equation}
\begin{split}
\left\{\text{exp}\left(\frac{3i\pi}{2}\hat{S}_1^{(2)}\right),\text{exp}~ i\pi\left(\hat{S}_1^{(1)}-\frac{\hat{S}_1^{(2)}}{2
}\right)&,\text{exp}\left(\frac{3i\pi}{2}\hat{S}_2^{(2)}\right),\text{exp}~ i\pi\left(\hat{S}_2^{(1)}-\frac{\hat{S}_2^{(2)}}{2}\right),\text{exp}\left(\frac{3i\pi}{2}\hat{Q}^{(1)}_\text{tot}\right),~\exp{i\pi\left(\hat{Q}^{(1)}_\text{tot}-\frac{\hat{Q}^{(2)}_\text{tot}}{2}\right)} \right\} \\
\equiv&\left\{e^{i\pi \hat{S}_{1A}},~e^{i\pi \hat{S}_{1B}},~e^{i\pi \hat{S}_{2A}},~e^{i\pi \hat{S}_{2B}}~e^{i\pi \hat{Q}_{totA}},~e^{i\pi \hat{Q}_{totB}} \right\}
\end{split}
\end{equation}
\fi
which have the eigenvalues
\begin{equation}
\begin{split}
\Big\{ e^{\frac{3i\pi}{2}s_{1}^{(2)}}, ~&e^{i\pi(s_1^{(1)}-s_1^{(2)})/3},~e^{\frac{3i\pi}{2}s_{2}^{(2)}},~e^{i\pi(s_2^{(1)}-s_2^{(2)})/3},~e^{\frac{3i\pi}{2} q^{(2)}_\text{tot}},~e^{i\pi(q_\text{tot}^{(1)}-q_\text{tot}^{(2)})/3} \Big\} \\ 
\equiv &\left\{ e^{i\pi s_{1A}},~e^{i\pi s_{1B}},~e^{i\pi s_{2A}},~e^{i\pi s_{2B}},~e^{i\pi q_{totA}},~e^{i\pi q_{totB}} \right\}
\end{split}    
\end{equation}
In this basis, by construction, the first entry $s_{1A}$ gives the number of $\psi_A$-like spin on $FM_1$ and the second entry $s_{1B}$ gives the number of $\psi_B$-like spins on $FM_1$ (and likewise, the third and fourth entries, $s_{2A}$ and $s_{2B}$ for $FM_2$.) Similarly, $q_{totA}$ gives the total charge in the system due $\psi_A$-type electrons and $q_{totB}$ that due to $\psi_B$-type electrons. The eigenstates are then denoted by the eigenvalues of the above operators
\begin{equation}
    \ket{s_{1A}, ~s_{1B},~s_{1A}, ~s_{1B},~q_{totA}, ~q_{totB} }
\end{equation}
which counts explicitly the number of different types of electrons in the system. From the allowed values of $s_j^{\alpha}$ and $q_\text{tot}^{(\alpha)}$, we can show the allowed values of spins and charges in the new basis to be $s_{jA/B},~q_{totA/B} \in \{0,~1\}$. We can also show that in the new basis, the charge parity of both types of electrons must be the same as the spin parity of the respective types of electrons. The action of charge operators on these states is as follows
\begin{equation}
\begin{split}
e^{i\pi \hat{Q}_1}  \ket{s_{1A}, ~s_{1B},~s_{1A}, ~s_{1B},~q_{totA}, ~q_{totB} } &=  \ket{s_{1A}-1, ~s_{1B},~s_{1A}+1, ~s_{1B},~q_{totA}, ~q_{totB} }\\
e^{3i\pi \hat{Q}_1^{(1)}}  \ket{s_{1A}, ~s_{1B},~s_{1A}, ~s_{1B},~q_{totA}, ~q_{totB} } &=  \ket{s_{1A}, ~s_{1B}-1,~s_{1A}, ~s_{1B}+1,~q_{totA}, ~q_{totB} }
\end{split}    
\end{equation}
where $e^{i\pi \hat{Q}_j}$ transfers a $\psi_A-$type electronic spin from $FM_j$ to $FM_{j+1}$, and $e^{3i\pi \hat{Q}_j^{(1)}}$ transfers a $\psi_B-$type electronic spin from $FM_j$ to $FM_{j+1}$. These transfers are explicit in the new basis. This is the basis we use to enumerate the ground-state manifold in Table.\ref{tab:Table1}.

\section{Interface Operators}

We now define physical operators defined at the interface of the SC/FM interface which is analogous to the Majorana operators for $\nu=1$ (with clean and robust $\nu=1$ edge) case which will be useful for the topological manipulation in low-energy subspace. We define the operator $\chi_{2i-1,\sigma}$, at the interface of $SC_{i}$ and $FM_{i}$ and $\chi_{2i}$ at the interface of $FM_{i}$ and $SC_{i+1}$ (see fig.\ref{fig:para_ed_recon}), with $SC_{3}$ identified as $SC_{1}$. Then,
\begin{eqnarray}
    \chi_{1\sigma} &=& e^{i\sigma (\theta^{(1)}_{1} + \theta^{(1)}_{2})} \hat{T}^{C}_{1}\hat{T}^{C}_{2} \nonumber\\
    \chi_{2\sigma} &=& e^{i\sigma (\theta^{(1)}_{1} + \theta^{(1)}_{2})} \hat{T}^{C}_{1}\hat{T}^{C}_{2}e^{i\pi\hat{S}_{1}} \nonumber\\
    \chi_{3\sigma} &=& e^{i\sigma (\theta^{(2)}_{1} + \theta^{(2)}_{2})} \hat{T}^{C}_{1}\hat{T}^{C}_{2}e^{i\pi\hat{S}_{1}} \nonumber\\
    \chi_{4\sigma} &=& e^{i\sigma (\theta^{(2)}_{1} + \theta^{(2)}_{2})} \hat{T}^{C}_{1}\hat{T}^{C}_{2}e^{i\pi\hat{S}_{tot}} 
\end{eqnarray}
where $e^{i\sigma(\theta^{(i)}_{1} + \theta^{(i)}_{2}))}$ adds a spin of $\sigma$ (consisting of $\sigma/3$ and $(2/3)\sigma$) in the $i^{th}$ FM region, and $\hat{T}^{C}_{1}\hat{T}^{C}_{2}$ increases the total charge by $e$ (consisting of $e/3$ and $(2/3)e$). As such the operator $\chi_{i\sigma}$ adds a charge and spin of e and $\sigma$ (of the type $1/3 + 2/3$), respectively. They follow the anti-commutation relation given by $\lbrace \chi_{i\sigma},\chi_{j\sigma'}\rbrace=0 $ for $i\neq j$. Also, $[\chi_{i\sigma},\chi_{i,\sigma'}]=0$. 
\begin{center}
\begin{figure}
        \centering
        \includegraphics[scale=0.6]{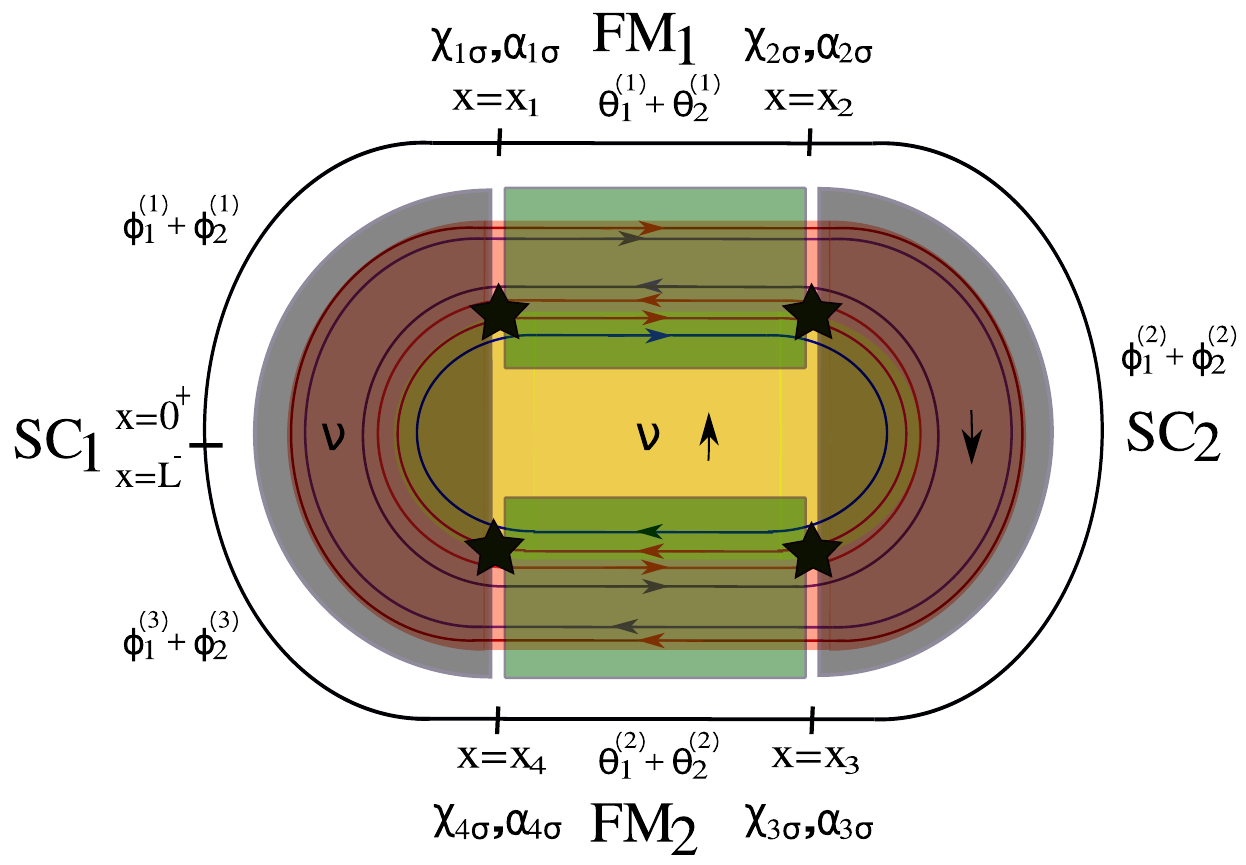}
        \caption{Schematic shows two quantum Hall systems (shown in light red and light yellow) with different spin ($\uparrow$ and $\downarrow$) and same filling fraction $\nu=1$. The boundary of QH systems (of length $L$) has reconstructed edge states with $1/3$ (in red) and $2/3$ (in blue) conductance. The edge states are gapped out by proximitized superconductors ($SC_{1}$ and $SC_{2}$) and ferromagnets ($FM_{1}$ and $FM_{2}$) alternatively. The coordinates are chosen such that the interface between SC and FM are at $x_{1}$, $x_{2}$, $x_{3}$, and $x_{4}$. $\phi^{j}_{i}/\theta^{j}_{i}$ denotes pinned $i^{th}$ bosonic fields pinned at the $j^{th}$ SC/FM regions, where $i=1$ and $i=2$ denotes the $1/3$ and the $2/3$ edge respectively. Zero modes localized at four interfaces are shown by the star. The "Majorana" zero modes, $\chi_{i\sigma}$ and $\alpha_{i\sigma}$ are localized at the interface at $x_{i}$. }   \label{fig:para_ed_recon}
    \end{figure}
\end{center}
Similarly, another set of such interface operators could be defined which solely consists of $1/3$ quasi-particles, such that
\begin{eqnarray}
    \alpha_{1\sigma} &=& e^{3i\sigma \theta^{(1)}_{1}} (\hat{T}^{C}_{1})^{3} \nonumber\\
    \alpha_{2\sigma} &=& e^{3i\sigma \theta^{(1)}_{1}} (\hat{T}^{C}_{1})^{3}e^{3i\pi\hat{S}^{(1)}_{1}} \nonumber\\
    \alpha_{3\sigma} &=& e^{3i\sigma \theta^{(2)}_{1}} (\hat{T}^{C}_{1})^{3}e^{3i\pi\hat{S}^{(1)}_{1}} \nonumber\\
    \alpha_{4\sigma} &=& e^{3i\sigma \theta^{(2)}_{1}} (\hat{T}^{C}_{1})^{3}e^{3i\pi\hat{S}^{(1)}_{tot}}, 
\end{eqnarray}
where $S^{(1)}_{1}$ is the spin content of the $1/3$ edge in the $FM_{1}$ region and $S^{(1)}_{tot}$ is $1/3$ edge contribution to the total spin content of the system. Similar to the previous $\chi_{i\sigma}$ operators, $\alpha_{i,\sigma}$ also creates a charge and spin of $e$ and $\sigma$ (three quasi-particles of 1/3), respectively. The operators $\alpha_{i\sigma}$ follow the commutation relation given by $\lbrace \alpha_{i\sigma},\alpha_{j\sigma'} \rbrace = 0$ for $i\neq j$ and $[\alpha_{i\sigma},\alpha_{i\sigma'}] = 0$. The commutation relation between $\chi_{i\sigma}$ and $\alpha_{j\sigma'}$ is given by $\lbrace \chi_{i\sigma},\alpha_{j\sigma'} \rbrace=0$ for $i\neq j$ and $[\chi_{i\sigma},\alpha_{i\sigma'}]=0$.

The action of operators $\chi_{i\sigma}$ and $\alpha_{i\sigma}$ on the states in the ground state manifold is given by
\begin{eqnarray}
    \chi_{1\sigma}|s_{1A},s_{1B};s_{2A},s_{2B};q_{totA},q_{totB} \rangle &=& e^{i\sigma (\theta^{(1)}_{1} + \theta^{(1)}_{2})} \hat{T}^{C}_{1}\hat{T}^{C}_{2} |s_{1A},s_{1B};s_{2A},s_{2B};q_{totA},q_{totB} \rangle \nonumber\\
    &=& |s_{1A}+\sigma,s_{1B};s_{2A},s_{2B};q_{totA}+1,q_{totB} \rangle \nonumber\\
    \chi_{2\sigma}|s_{1A},s_{1B};s_{2A},s_{2B};q_{totA},q_{totB} \rangle &=& e^{i\sigma (\theta^{(1)}_{1} + \theta^{(1)}_{2})} \hat{T}^{C}_{1}\hat{T}^{C}_{2} e^{i\pi \hat{S}_{1}}|s_{1A},s_{1B};s_{2A},s_{2B};q_{totA},q_{totB} \rangle \nonumber\\
    &=& e^{i\pi s_{1}}|s_{1A}+\sigma,s_{1B};s_{2A},s_{2B};q_{totA}+1,q_{totB} \rangle \nonumber\\
    \chi_{3\sigma}|s_{1A},s_{1B};s_{2A},s_{2B};q_{totA},q_{totB} \rangle &=& e^{i\sigma (\theta^{(2)}_{1} + \theta^{(2)}_{2})} \hat{T}^{C}_{1}\hat{T}^{C}_{2} e^{i\pi \hat{S}_{1}}|s_{1A},s_{1B};s_{2A},s_{2B};q_{totA},q_{totB} \rangle \nonumber\\
    &=& e^{i\pi s_{1}}|s_{1A},s_{1B};s_{2A}+\sigma,s_{2B};q_{totA}+1,q_{totB} \rangle \nonumber\\
    \chi_{4\sigma}|s_{1A},s_{1B};s_{2A},s_{2B};q_{totA},q_{totB} \rangle &=& e^{i\sigma (\theta^{(2)}_{1} + \theta^{(2)}_{2})} \hat{T}^{C}_{1}\hat{T}^{C}_{2} e^{i\pi \hat{S}_{tot}}|s_{1A},s_{1B};s_{2A},s_{2B};q_{totA},q_{totB} \rangle \nonumber\\
    &=& e^{i\pi s_{tot}}|s_{1A},s_{1B};s_{2A}+\sigma,s_{2B};q_{totA}+1,q_{totB} \rangle
\end{eqnarray}

\begin{eqnarray}
    \alpha_{1\sigma}|s_{1A},s_{1B};s_{2A},s_{2B};q_{totA},q_{totB} \rangle &=& e^{3i\sigma \theta^{(1)}_{1}} (\hat{T}^{C}_{1})^{3} |s_{1A},s_{1B};s_{2A},s_{2B};q_{totA},q_{totB} \rangle\nonumber\\
    &=& |s_{1A},s_{1B}+\sigma;s_{2A},s_{2B};q_{totA},q_{totB}+1 \rangle \nonumber\\
    \alpha_{2\sigma}|s_{1A},s_{1B};s_{2A},s_{2B};q_{totA},q_{totB} \rangle &=& e^{3i\sigma \theta^{(1)}_{1}} (\hat{T}^{C}_{1})^{3} e^{3i\pi \hat{S}^{(1)}_{1}}|s_{1A},s_{1B};s_{2A},s_{2B};q_{totA},q_{totB} \rangle \nonumber\\
    &=& e^{3i\pi s^{(1)}_{1}}|s_{1A},s_{1B}+\sigma;s_{2A},s_{2B};q_{totA},q_{totB}+1 \rangle \nonumber\\
    \alpha_{3\sigma}|s_{1A},s_{1B};s_{2A},s_{2B};q_{totA},q_{totB} \rangle &=& e^{3i\sigma \theta^{(2)}_{1}} (\hat{T}^{C}_{1})^{3} e^{3i\pi \hat{S}^{(1)}_{1}}|s_{1A},s_{1B};s_{2A},s_{2B};q_{totA},q_{totB} \rangle \nonumber\\
    &=& e^{3i\pi s^{(1)}_{1}}|s_{1A},s_{1B};s_{2A},s_{2B}+\sigma;q_{totA},q_{totB}+1 \rangle \nonumber\\
    \alpha_{4\sigma}|s_{1A},s_{1B};s_{2A},s_{2B};q_{totA},q_{totB} \rangle &=& e^{3i\sigma \theta^{(2)}_{1}} (\hat{T}^{C}_{1})^{3}e^{3i\pi \hat{S}^{(1)}_{tot}}|s_{1A},s_{1B};s_{2A},s_{2B};q_{totA},q_{totB} \rangle \nonumber\\
    &=& e^{3i\pi s^{(1)}_{tot}}|s_{1A},s_{1B};s_{2A},s_{2B}+\sigma;q_{totA},q_{totB} +1\rangle
\end{eqnarray}
Also, $\chi_{i\sigma}^{2} |s_{1A},s_{1B};s_{2A},s_{2B};q_{totA},q_{totB} \rangle = |s_{1A},s_{1B};s_{2A},s_{2B};q_{totA},q_{totB} \rangle$ and $\alpha_{i\sigma}^{2} |s_{1A},s_{1B};s_{2A},s_{2B};q_{totA},q_{totB} \rangle = |s_{1A},s_{1B};s_{2A},s_{2B};q_{totA},q_{totB} \rangle$. Hence, $\chi_{i\sigma}$ and $\alpha_{i\sigma}$ operators are the zero modes at the SC-FM interface are the realization of Majorana zero modes in our system.

\section{Josephson Junction}
The superconductors on either side of the Josephson junction are modelled by the Hamiltonian
\begin{equation}
\begin{split}
    H_{SC} = -\sum_{\gamma=A,B,C} \Delta_\gamma \left(\int_{ SC_1} dx~ \psi_\gamma \psi_\gamma +\int_{SC_2 }dx~  e^{i\phi_{SC}} \psi_\gamma \psi_\gamma + hc \right)
    \label{Eq:JJ_pairing} 
\end{split}    
\end{equation}
which in the bosonized form reads
\begin{equation}
\begin{split}
&H_{SC} = - \int_{SC_1} dx~ \left[ \Delta_A\cos{(2(\phi_1 +\phi_2))} + \Delta_B \cos{(6\phi_1)} + \Delta_C \cos{(3\phi_2)} \right]   \\ &- \int_{SC_2} dx~ \left[ \Delta_A\cos{(2(\phi_1 +\phi_2) + \phi_{SC})} + \Delta_B \cos{(6\phi_1+ \phi_{SC})} + \Delta_C \cos{(3\phi_2+ \phi_{SC})} \right]     
\end{split}    
\end{equation}
where the superconducting phase has been plugged into the second superconductor. In the limit $Delta_\gamma \rightarrow \infty$, these cosines are all pinned to one of their minima, resulting the in the following boundary conditions of the bosonic fields
\begin{eqnarray}
  (\varphi_1 + \varphi_2)(L_1) &=& 0;~ \varphi_{1}(L_1) = 0;~ \varphi_{2}(L_1) = 0 \nonumber\\
  (\varphi_1 + \varphi_2)(L_2) &=& \hat{F};~ \varphi_{1}(L_2) = \hat{F}_1;~ \varphi_{2}(L_2) = \hat{F}_2 
    \label{boundary_conditions}
\end{eqnarray}
\begin{gather}
   \hat{F} = \mathrm{mod}\left[\pi\left(\frac{\hat{N}^{\varphi_1}_{2}}{3} + \frac{2\hat{N}^{\varphi_2}_{2}}{3} - \frac{\phi_{SC}}{2\pi}\right) + \pi,2\pi\right] - \pi \nonumber\\
    \hat{F}_{2} = \frac{2}{3}\left( \mathrm{mod}\left[ \pi \left(\hat{N}^{\varphi_2}_{2}-\frac{\phi_{SC}}{2\pi}\right)+\pi,2\pi\right]-\pi\right)\nonumber\\
    \hat{F}_1 = \hat{F}-\hat{F}_2
    \label{compactness}
\end{gather}
where the mods are imposed to ensure single-valuedness of the electron operators, and spins $s_{1}^{(1)}$ and $s_{1}^{(2)}$ to be consistent with the state of the system. The effective Hamiltonian of the Josephson junction is given by the free bosonic Hamiltonian
\begin{eqnarray}
    H_{\text{eff}} &=& \sum_{\alpha=1}^{2}\frac{\hbar v_\alpha}{2\pi\nu_\alpha}\int_{L_1}^{L_2} dx \left[ \left(\partial_{x} \varphi_{\alpha}\right)^{2} +  \left(\partial_{x} \theta_{\alpha}\right)^{2} \right]
\end{eqnarray}
where the bosonic fields $\varphi_{j}$ and $\theta_{j}$ are subject to the boundary conditions of Eqns. \ref{boundary_conditions} and \ref{compactness}. The bosonic field expansion that satisfies these boundary conditions and diagonalizes the Hamiltonian is given by
\begin{eqnarray}
    \varphi_{\alpha}(x) &=& \hat{\varphi}_{\alpha}(L_2)\frac{(x-L_1)}{L_J} +\sqrt{\nu_{\alpha}}\sum_{k=1}^{\infty}i \frac{\sin\lambda_{k}(x)}{\sqrt{k}}\left(\hat{a}_{\alpha,k} - \hat{a}^{\dagger}_{\alpha,k}\right) \nonumber\\
    \theta_{\alpha}(x) &=& \theta^{(0)}_\alpha + \sqrt{\nu_{\alpha}}\sum_{k=1}^{\infty} \frac{\cos\lambda_{k}(x)}{\sqrt{k}}\left(\hat{a}_{\alpha,k} + \hat{a}^{\dagger}_{\alpha,k}\right)
    \label{Eq:Field_expansion}
\end{eqnarray}
where, $\hat{a}_{\alpha,k}$ is the bosonic annihilation operator for the $\alpha^{th}$ bosonic mode, with commutation relations $\left[\hat{a}_{\alpha,k},\hat{a}^{\dagger}_{\beta,k'}\right]=\delta_{kk'}\delta_{\alpha\beta}$, $\lambda_{k}(x) = \frac{k\pi}{L_J}(x-L_1 )$, and $L_J = L_2-L_1$. $\hat{\varphi}_{\alpha}(L_2)$ and $\theta^{(0)}_\alpha$ are zero-mode operators. Correct commutation relations between the $\varphi, \theta$ fields is ensured by imposing $\left[\hat{N}^{\varphi_\alpha}_{2},\hat{\theta}^{(\alpha)}_{0}\right] = i$. Plugging the field expansion of Eq.\ref{Eq:Field_expansion}, into the effective Hamiltonian
\begin{eqnarray}
    H_{\text{eff}} &=& \sum_{\alpha=1}^{2}\hbar v_{\alpha} \left[ \frac{\left(\hat{\varphi}_{\alpha}(L_{2})\right)^{2}}{2\nu_{\alpha}\pi L_J} + \sum_{k=1}^{\infty} \frac{\pi k}{L_{J}}\left(\hat{a}^{\dagger}_{\alpha,k}\hat{a}_{\alpha,k} + \frac{1}{2}\right) \right]\nonumber\\
    \label{Eq:hamiltonian_2}
\end{eqnarray} 
from which the Josephson current can be readily obtained as the derivative of the Hamiltonian with respect to the superconducting phase.}

\end{document}